# Reflectance of Rhenium as a Function of Pressure in a Diamond Anvil Cell

Jing Song*[1], Irina Chuvashova*[2], and Isaac F. Silvera**

Lyman Laboratory of Physics, Harvard University, Cambridge MA, 02138

**Abstract**

We have measured the reflectance of rhenium in the visible region to pressures up to 100 GPa in a diamond anvil cell (DAC). By photographing the reflecting surface, we visually show that there are challenges to obtaining accurate values in a DAC for several reasons and propose some useful procedures. We also show that knowledge of the reflectance of rhenium can overcome the problem of absorption of light by diamonds when studying the reflectance of materials at high pressure in a DAC.

1. Current address: Institute of Physics, Chinese Academy of Sciences, Beijing 100190, China.
2. Current address: Dept. of Chemistry and Biochemistry, Florida International University, 11200 SW 8t St, Miami, Fl 33199.
 *Equal contributions.
**Corresponding author. Email:  Silvera@physics.harvard.edu

There are two standard experimental methods to determine that a material is a metal: measurement of electrical conductivity or measurement of reflectance, R, both as a function of temperature. For a metal the conductivity remains finite in the limit that temperature goes to zero Kelvin, while the reflectance, due to free electrons, remains relatively independent of temperature. A proper measurement of conductance requires attaching 4 wires to a sample which is challenging for very small samples. The measurement of reflectance is generally a simpler procedure: a collimated light beam of intensity $I_0$ (we call this the reference beam) is reflected off the sample surface with intensity $I_R$. Both intensities are measured and the reflectance $R= I_R/I_0$ at a given angle and polarization.

In recent decades the study of insulator to metal transitions (IMTs) has been driven both by theoretical predictions, and by high-pressure to increase particle density in studies of samples. The most prominent theoretical example is the 1935 prediction by Wigner and Huntington [1] that under pressure, solid molecular hydrogen would undergo an IMT to an atomic solid. The original predicted transition pressure was 25 GPa, while modern calculations are ~20 times higher in the 400-500 GPa region. The transition to metallic hydrogen has recently been observed in reflection measurements in a diamond anvil cell (DAC) at a pressure of 495 GPa [2].

High pressures can be achieved statically in DACs or dynamically by shock compression; the former has the advantage that very low temperatures can be achieved. In this paper we focus on DACs. The heart of a DAC is shown in Fig. 1 where a sample is confined in a gasket and compressed between the culet flats of the two diamonds. To achieve multi-megabar pressures, culet-flat dimensions are very small, of the order of 30 microns or less. Thus, it is challenging to insert wires into the sample to measure the conductivity and multiple attempts may be required to succeed, while reflectance measurements are easier to prepare, but have some challenges.

In order to measure R in a DAC one generally encounters two problems. First, the intensity of the reference beam $I_0$ must be measured. One procedure for a DAC in air (or vacuum) is to measure the reflectance off the diamond table $I_D=I_0 R_{D-air}$, and knowing the index of refraction of diamond, one can calculate to find $I_0$, using $R_{D-air}=(n_{diam}-n_{air})^2/(n_{diam}+n_{air})^2$. Second, if the diamond has absorption in the spectral region of measurement, the absorption will attenuate the reflected beam by an unknown amount, thus excluding measurements in this region.

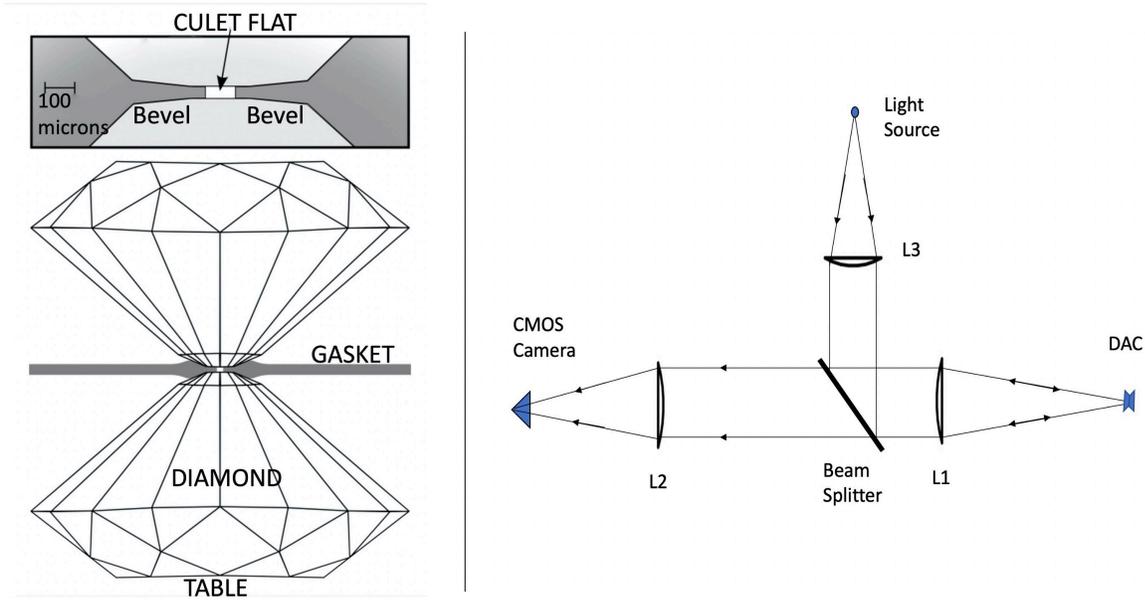

Figure 1. Left: a typical pair of diamonds compressing a gasket; diamond dimensions are a few millimeters. Upper left: an expansion of the heart of the DAC showing the culet flat and gasket. The gasket has a hole for placement of a sample; optical access is normal to the sample or the culet flat. Light enters the diamond at the table and focuses on the culet flat. Right: optical layout for measuring reflectance of sample in a diamond anvil cell.

In this paper we present a method to overcome the absorption problem. Most gaskets in DACs used for ultra-high pressure are made of high purity rhenium because it is very hard, and yet deformable. If the pressure dependence of the reflectance of rhenium, $R_{Re}$, is a known quantity, then the diamond absorption problem can be resolved and the reflectance of the sample, $R_{sample}$, can be determined. If one confines a sample in a Re gasket, one can measure the reflectance of both the sample and the Re and use the known reflectance of Re to determine the reference beam. Even if the diamond is absorbing with absorption coefficient $\alpha(\lambda)$, this cancels out, i.e.

$$I_{sample} = I_0 e^{-2\alpha(\lambda)d} * R_{sample} \tag{1a}$$

$$I_{Re} = I_0 e^{-2\alpha(\lambda)d} * R_{Re} \tag{1b}$$

so

$$\frac{I_{sample}}{I_{Re}} = \frac{R_{sample}}{R_{Re}} \tag{2a}$$

$$R_{sample} = R_{Re} * \frac{I_{sample}}{I_{Re}} \tag{2b}$$

The factor of 2 in Eq. 1 is due to the double pass of the light through the diamond. Thus, it is useful to know the experimental value of the rhenium gasket, $R_{Re}$. This is the subject of this paper.

In a DAC the geometry forces the angle of incidence to be approximately zero, or normal to the surface as seen in Fig.1, left. The light enters the DAC at the table of the diamond and even if the light is focused on the sample surface (see Fig. 1, right), most of the rays are almost normal to the surface due to Snell's law and the large index of refraction of diamond, around 2.3 to 2.4 in the visible/IR wavelength region. The light reflected off the sample first passes through a diamond and back out to a detector. The three-dimensional region of the diamond that contacts the sample (the culet flat) is highly stressed, while the opposite end, the table, is large (several square millimeters) and is essentially unstressed. The diamonds typically have an 8 to 9 degree bevel outside the culet flat, with a diameter of about 300 microns.

In the near and mid-infrared (IR), unstressed diamond has spectral regions of intrinsic absorption that are usually avoided [3]. However, under very high stress, (pressures of order 200-500 GPa), the diamonds can have absorption in the visible [4]; this can prevent the measurement of reflectance, as the absorption can arise from sources such as impurities in the diamond and $\alpha(\lambda,P)$ is unknown from diamond to diamond, as well as the stress distribution in the diamond. The stress is highest in the culet region and minimal near the table, so $\alpha(\lambda)$ depends on the position in the diamond and should be considered as an average or integral over the pathlength. Absorption by the diamond was a problem in the recent observation of metallic hydrogen at low temperature and ultra-high pressure (~500 GPa) [2,5]. In this paper we show a possible method of overcoming this diamond absorption problem by measuring the reflectance of rhenium; this was carried out to pressures of order 100 GPa.

We have measured the reflectance of Re in a DAC at room temperature using diamonds with 80-micron culet flats, shown in Fig. 1, left. The focal spot of the light source was about 100 microns and overfilled the culet of the diamond, so in principle the illumination of the sample should be uniform. The detector was a Thorlabs color CMOS camera (model DC1645C) with useful functional software for measurement of intensity and analysis. This camera has a Bayer filter [6] and can measure intensities in the red (610 nm), green (537 nm), and blue (460 nm) in a reasonably narrow band.

We indented a Re gasket and measured the reflectance of the Re pressed to the surface of the diamond culet. A hole in the gasket was filled with ruby in a quasi-hydrostatic pressure medium of KCl [7]. All measurements were made in a Vascomax DAC (identity VMII) [8]. To simplify the measurements, we placed a dot of silver (Ag) on the gasket indentation. Silver has a reflectance of greater than ~0.90-0.99 in the visible-IR and is expected to approach 1 with increasing pressure, as the density of carriers increases and the plasma frequency shifts further into the UV [9]. Thus, we can use the reflectance of the Ag for normalization of the reflectance of the Re, similar to the scheme discussed for Eqs. 1-2.

Surprisingly we encountered several problems that arise due to the measurements in a DAC, that resulted in large uncertainties. We prepared four samples by different methods and measured the reflectance. One of the diamonds used in these measurements had a small ring crack. This was not limiting and our pressure ranged up to ~100 GPa, but had other deleterious effects. Because of the use of Ag for normalization, we used the technique of measuring reflectance off the diamond table in only a few of the later runs.

For Run 1, we shall describe techniques in detail that are common to the four runs. The maximum pressure in this run was 105 GPa. Figure 2 shows the sample viewed with the CMOS color camera. Figure 3 sketches the configuration of rhenium, silver, and ruby between the diamond culets for the four runs. In Run 1 the Re gasket has a through hole in the center that was filled with ruby in a KCl pressure medium; at the bottom right, a dot of silver was flattened onto the gasket with the diamonds. Figure 2a shows an image of the sample region in reflected light. The highly reflective silver is seen at about 5 o'clock. The camera software enables one to plot the red, green, blue (RGB) intensity values along a line through the image (red line in the figure). One can also pick a point in the figure and get a 5x11 table of RGB values, surrounding the point; this enables us to find the value and the standard deviation of the measurement for each wavelength. We calculate reflectance ($R_{Re}$) as

$$R_{Re} = \frac{I_{Re} - I_{Black}}{I_{Ag} - I_{Black}} * R_{Ag} \qquad (3)$$

and the standard deviation σ, as

$$\sigma(R_{Re}) = \sqrt{\frac{\sigma(I_{Re})^2 + \sigma(I_{Black})^2}{(I_{Re} - I_{Black})^2} + \frac{\sigma(I_{Ag})^2 + \sigma(I_{Black})^2}{(I_{Ag} - I_{Black})^2}} * R_{Re} \,. \qquad (4)$$

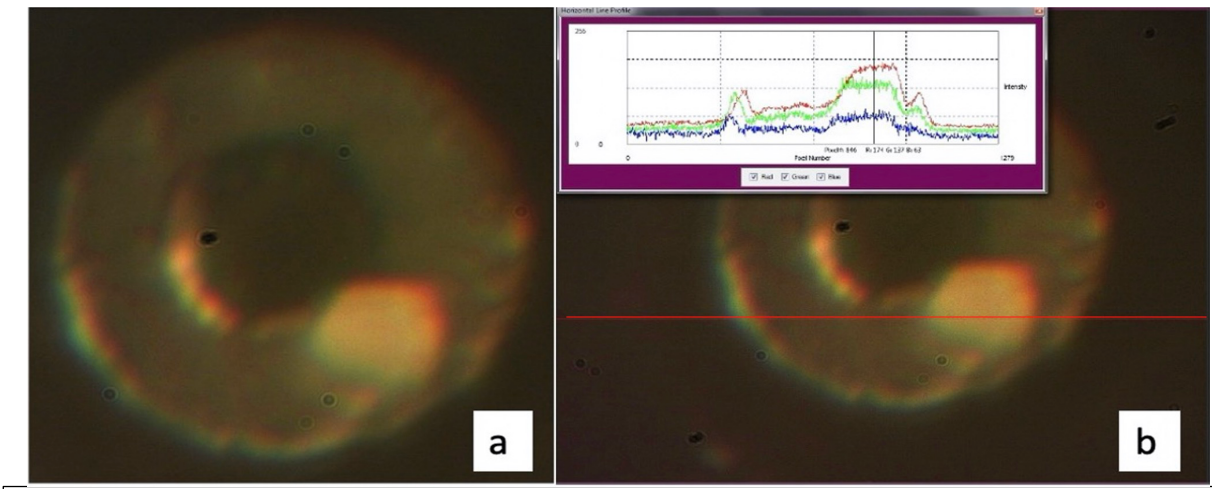

Fig. 2, Run 1, ambient pressure in the DAC. a) is a camera image of reflected light, and b): the inset shows the intensity of the reflected signal in red, green, blue (RGB) along the horizontal red line. The important things to notice are the relative uniform reflectance of the rhenium and the silver, the ruby/KCl (which appears black, as ruby and KCl are almost transparent), and the black level surrounding the Re flat. (The fonts in the inset in b are left undersized, as they are not important for this figure or those to follow.)

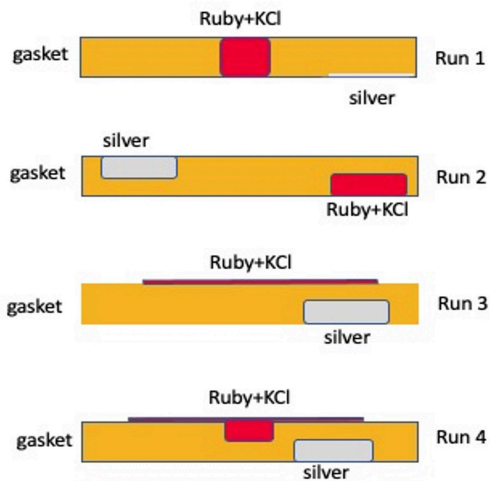

Fig. 3. The configuration of the gasket in the four runs. In each run the Re gasket had ruby mixed with KCl as a pressure medium, and silver as a reference reflector.

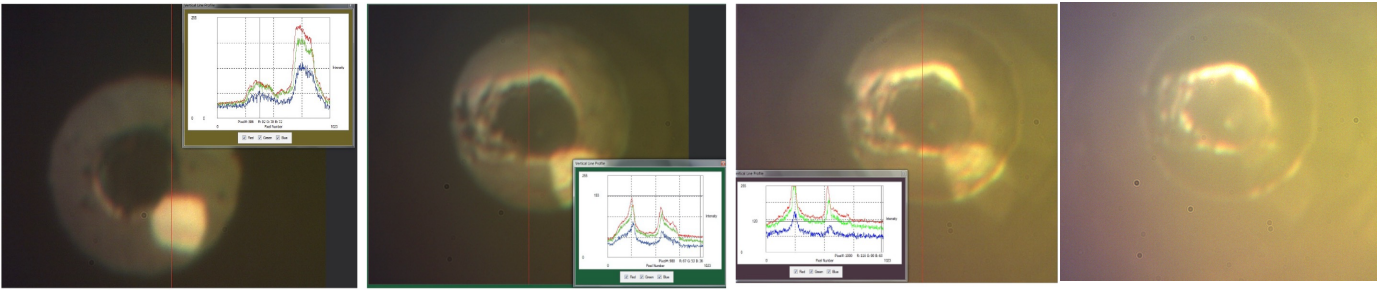

Fig. 4, Run1. From left to right: images in reflected light at pressures of 10, 48, 82, and 105 GPa.

Here, $I_{black}$ is the background or black level outside of the diamond flat; we consider the background level in different regions of the sample to be uncorrelated. We evaluated 3 tables, each from different points on the sample for every pressure point in order to determine the non-uniformity of the measured reflectance. Measurements of reflectance were made for up to 10 pressure values in each run; we show images of some of these in Fig. 4.

      There were a few limiting problems in Run 1. The silver is much softer than the rhenium, so as the pressure or load increased, the silver, which was originally relatively thick and flat on the rhenium, thinned and translated out of the culet region; it eventually became thinner than the optical penetration depth and the underlying Re could be seen. This was more extreme at 105 GPa shown in Fig. 4; this data could not be used. A second problem can be seen in the reflectance at 82 and 105 GPa in Fig. 4: due to scattered light the background level was not uniform and this led to a large uncertainty. The third, related, problem was cupping of the diamond culet [10]. At high pressure the culet region elastically deforms in a cup shape and reflectance depends on the angle of incidence which varies as the culet is no longer flat; the beveled region also deforms so that it is closer to the plane of the culet. Due to the overfilling of the light focused on the sample, light could scatter off the facets of the bevel, the pavilion of the diamond, and stress created defects. This cupping and the ring crack in the diamond led to enhanced scattering and a poorly defined black level, resulting in large uncertainties. A better example of the problem of cupping is shown in Run 4. The reflectance results of Run 1 are shown in Fig. 5.

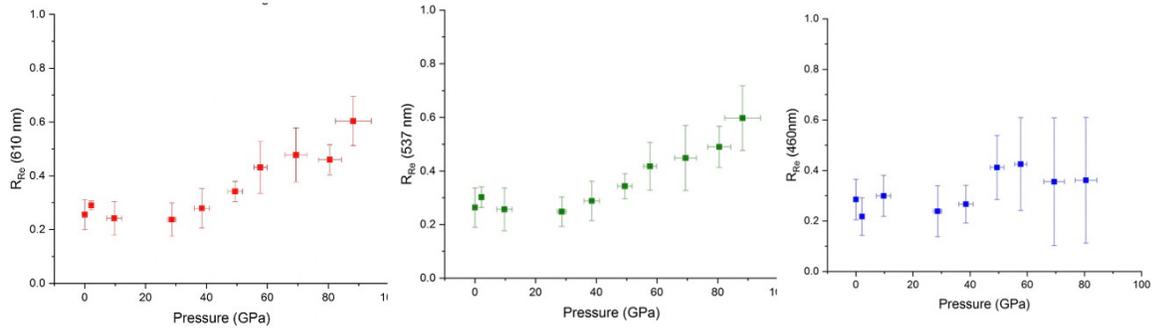

Fig. 5. Reflectance of Re in Run:1 red, green, blue (RGB) from left to right. The highest pressure data are shown (blue points, 460 nm), but from Fig. 4 it clear that the points for pressures above 80 GPa are of low quality, and higher pressure points up to 105 GPa are not shown. Data is based on 3 tables.

In Run 2, in order to resolve the problem of the thinning of the silver, we made a half hole in the gasket by electric discharge machining (EDM) and filled it with a clump of silver (see Fig. 3). Under load, some of the silver spread over the surface (see Fig. 6), but this resolved the problem of thinning of the silver. A similar half-hole was made for ruby on the other side of the gasket. The gasket was 30 microns thick before loading. As pressure increased the ruby/KCl sample translated radially out of the culet region and we could no longer reliably measure pressure; the maximum pressure of this run was 40.3 GPa.

In Run 3, we spread a ruby/KCl mixture on one side of the gasket; ruby particles were about 3 microns in dimension. The thickness of the gasket started at 20 microns and the maximum pressure reached was 62.3 GPa. Reflectance was measured on the side of the DAC with the diamond with no ring crack, to minimize scattered light. When pressure was increased the surface became rough and we observed non-uniform reflection of both silver and Re. This is an example of unexpected behavior in which the sample seems to roughen, shown in Fig. 7. This is probably due to the pressure gradient and the shear stress on the rhenium surface.

The final measurements were in Run 4. A central half-hole was EDM'd for the ruby/KCl with a smear of ruby/KCl on the surface, and a half-hole on the other side for the silver. A pressure of 91 GPa was achieved. The starting gasket thickness was just under 20 microns. The limiting problem was that the clump of silver was too close to the edge of the culet and eventually translated radially out of view as seen in Fig 8; at a pressure of 91 GPa no silver is seen.  The 91 GPa image also shows that the reflectance was very uneven due to cupping of the diamonds and there was a strong intensity gradient of reflected light from the center of the culet to the edge.  After we observed the silver to become problematic, we measured reflectance off the table of the diamond in some of the pressure points, and this data agrees fairly well with other data.

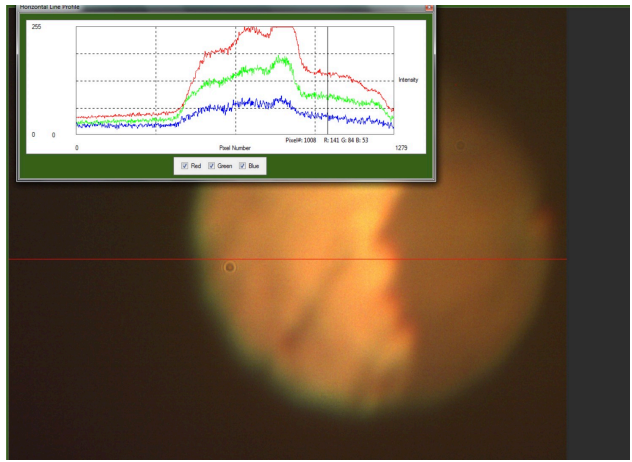

Fig. 6. Run 2 at 40 GPa. Again, the inset shows the intensity along the red line.

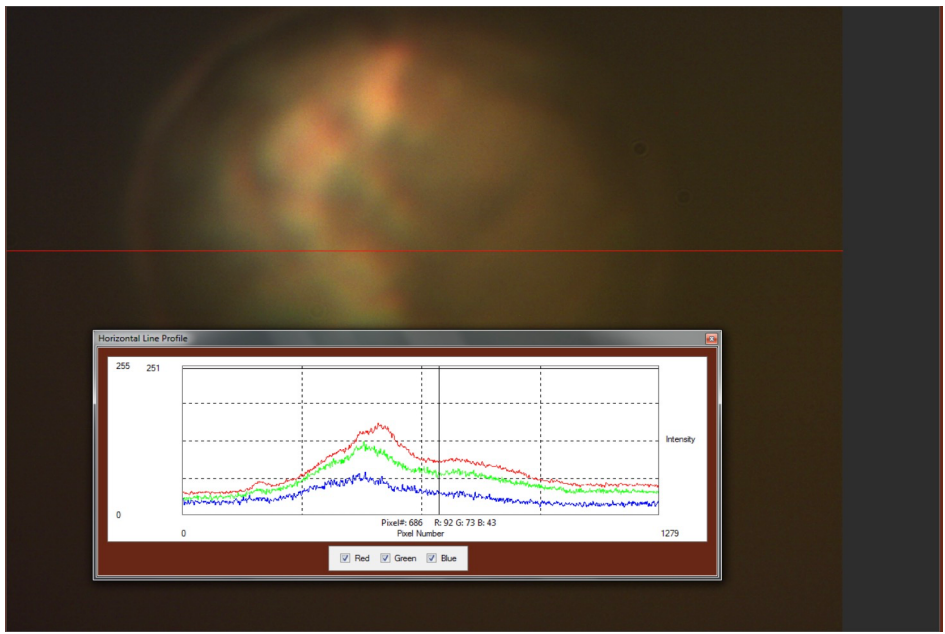

Fig. 7. Image of the culet region in Run 3 at a pressure of 62.3 GPa.

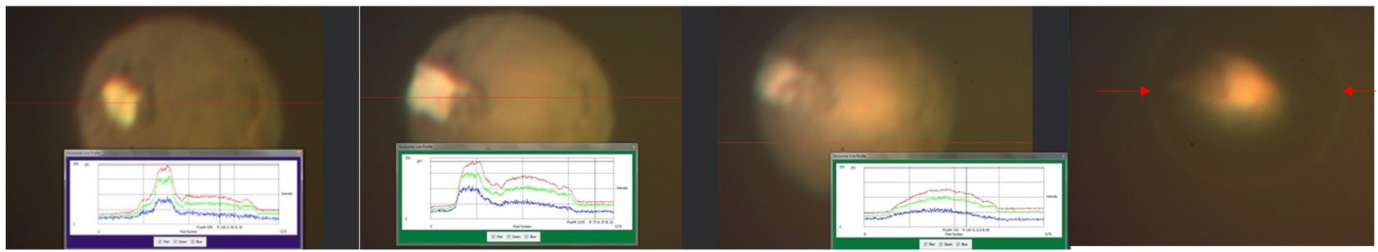

Fig. 8, Run 4. Pressure from left to right: 14, 36, 47, 91 GPa. The outer edge of the culet is the same for all four pressures, but this edge is difficult to see for 91 GPa due to cupping of the diamond. In this case, the red arrows indicate the edge of the culet flat; at this pressure the silver is no longer under the flat.

In our experiments several problems were encountered for measurement of reflectance. In the initial optical alignment, it is important for the light intensity to be uniform and normal to the ambient pressure surface of the sample. An encountered problem is that as pressure is increased, the diamond culet flats cup or deform, so that light is only normal to the surface at the center. From the reflectance traces across the sample, we estimate that regions of uniformity are at most around 20 microns for pressures greater than around 40-50 GPa. Radial spread of the sample (Re, Ag, and ruby) was a problem. This can be hindered by starting with much thinner gaskets; ours were initially about 20 to 30 microns thick. We observed unexpected development of surface roughness with mounting pressure. We have also seen this for soft samples, such as a thin layer of gold (50-100 nm thick) plated on rhenium. With mounting pressure, the gold tends to clump together and roughen rather than presenting a smooth uniform surface. We found that thin layers of Au do not tend to clump. Roughening was unexpected for a rhenium surface which is very hard, yet we observed roughening of the surface in some cases. Scattered light can be a problem in giving a non-uniform black level which adds to the uncertainty of the measurement. Thin samples should not suffer much from cupping; in the limit that the thickness goes to zero, the deformation of the culet flat should become negligible.

In Fig. 9 we summarize our results. We also show a result at ambient conditions from the book by Palik [11]. We provide a best fit to our data with a straight-line function. The fits are

$$\text{Red:} \quad R_{Re} = 0.295\,(19) + 1.88(59) * 10^{-3} * P \tag{5}$$

$$\text{Green:} \quad R_{Re} = 0.278\,(14) + 1.95(47) * 10^{-3} * P \tag{6}$$

$$\text{Blue:} \quad R_{Re} = 0.208\,(20) + 7.33(6.6) * 10^{-4} * P . \tag{7}$$

Values in parenthesis are uncertainties.

In this paper we have demonstrated that knowledge of the reflectance of rhenium can overcome the problem of absorption of light by diamonds when studying the reflectance of materials at high pressure in a DAC. Measurement of reflectance in a DAC requires great. Better results are obtained if the samples are small in area and well centered on the culet flat so that diamond cupping is not a problem. Gaskets should be thin to minimize cupping. Conditions are best if the sample completely fills the hole in the gasket such as a condensable gas, or the sample is in a quasi-hydrostatic pressurization media to minimize pressure gradients across the sample.

We thank the NSF, Grant No. DMR-1905943, and the DoE Stockpile Stewardship Academic Alliance Program, Grant No. DE-NA0003917, for support of this research. We also thank R. Dias and R. Vincent for useful discussions.

Data supporting our observations are available from the corresponding author upon reasonable request.

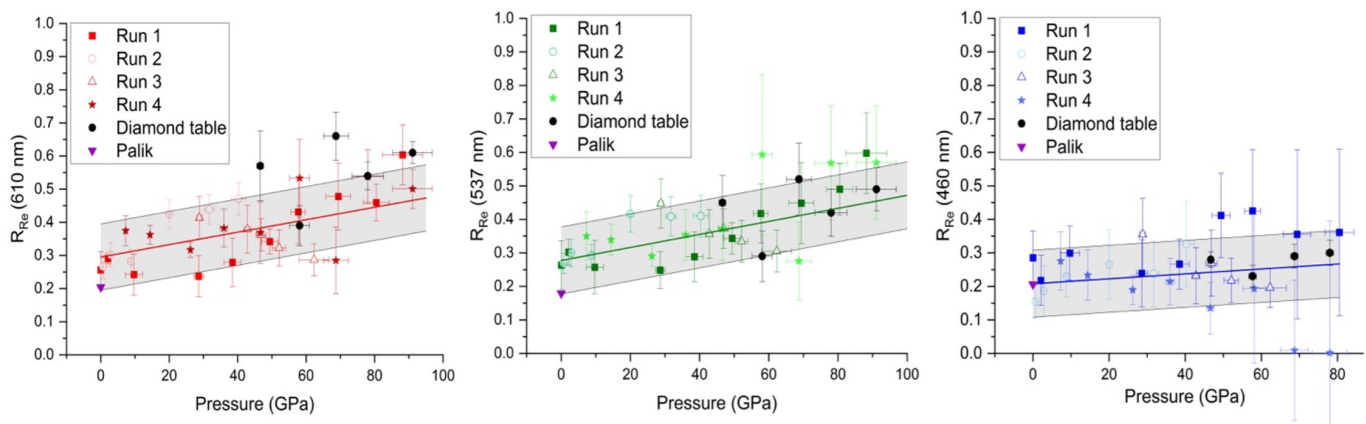

Figure 9. A summary of reflectance results for all runs using 3 tables. We also show normalization data due to reflectance from the diamond tables. The solid lines are fits of our data to a straight line, with the grey area as the one-$\sigma$ confidence interval.

**References**

[1] E. Wigner and H. B. Huntington, J. Chem. Phys. **3**, 764 (1935).

[2] R. Dias and I. F. Silvera, Science **355**, 715 (2017).

[3] A. M. Bennett, B. J. Wickham, H. K. Dhillon, Y. Chen, S. Webster, G. Turri, and M. Bass, Proceedings of Spie **8959** (2014).

[4] Y. K. Vohra, in *Proceedings of the XIII AIRAPT International Conference on High Pressure Science and Technology* (1991, Bangalore, India, 1991).

[5] I. F. Silvera and R. Dias, Science **357, eaao5843** (2017).

[6] Bayer Filter, (Wikipedia).

[7] A. Dewaele, A. B. Belonoshko, G. Garbarino, F. Occelli, P. Bouvier, M. Hanfland, and M. Mezouar, Phys. Rev. B **85**, 214105 (2012).

[8] I. F. Silvera and R. J. Wijngaarden, Rev. Sci. Instrum. **56**, 121 (1985).

[9] J. M. Bennett and E. J. Ashley, Applied Optics **4**, 221 (1965).

[10] R. J. Hemley, H.-k. Mao, G. Shen, J. Badro, P. Gillet, M. Hanfland, and D. Hausermann, Science **276**, 1242 (1997).

[11] E. D. Palik, *Handbook of Optical Constants of Solids* 1997), Vol. III.